\documentclass[aps,prc,twocolumn,showkeys,showpacs]{revtex4}
\usepackage{amssymb}
\usepackage{graphicx}
\usepackage{color}

\newcommand{\req}[1]{(\ref{#1})}
\newcommand{\bel}[1]{\begin{equation}\label{#1}}
\newcommand{\belar}[1]{\begin{eqnarray}\label{#1}}
\def\epsk{\epsilon_k}
\def\mev{\;{\rm MeV}}

\def\pr{\prime}

\begin{document}

\title{The temperature dependence of the shell corrections}

\author{F.A. Ivanyuk}
\email{ivanyuk@kinr.kiev.ua}
\affiliation{Institute for Nuclear Research, Prospect Nauki 47, 03028 Kiev, Ukraine}
\affiliation{Laboratory for Advanced Nuclear Energy, Institute of Innovative Research, Tokyo Institute of Technology, Tokyo, 152-8550 Japan}
\author{C. Ishizuka}
\email{chikako@nr.titech.ac.jp}
\affiliation{Laboratory for Advanced Nuclear Energy, Institute of Innovative Research, Tokyo Institute of Technology, Tokyo, 152-8550 Japan}
\author{M.D. Usang}
\email{mark_dennis@nuclearmalaysia.gov.my}
\affiliation{Laboratory for Advanced Nuclear Energy, Institute of Innovative Research, Tokyo Institute of Technology, Tokyo, 152-8550 Japan}
\affiliation{Malaysia Nuclear Agency, Bangi, Malaysia}
\author{S. Chiba}
\email{chiba.satoshi@nr.titech.ac.jp}
\affiliation{Laboratory for Advanced Nuclear Energy, Institute of Innovative Research, Tokyo Institute of Technology, Tokyo, 152-8550 Japan}
\affiliation{National Astronomical Observatory of Japan, Tokyo, Japan}
\date{today}

\begin{abstract}
We have examined the dependence of the shell correction to the nuclear liquid drop energy at finite excitations on the excitation energy (temperature). For this we have calculated the shell correction to the energy and free energy in very broad region of nuclei and deformations starting directly from their formal definitions. We have found out that the dependence of the shell corrections on the excitation energy differ substantially from the widely used approximation $\delta E(E^*)=\delta E(0)\exp(-E^*/E_d)$ both at small and large excitations. In particular, below the critical temperature at which the pairing effects vanish, the shell correction to the free energy is rather insensitive to the excitation energy.

We suggest a more accurate approximation for the temperature dependence of the shell correction to the energy and free energy that is expressed in terms of the shell correction to the energy of independent particles and the shell correction to the pairing energy at $T=0$ and few fitted constants.
\end{abstract}

\pacs{21.60.-n, 21.60.Cs, 25.85.Ec}
\keywords{shell correction, excitation energy, free energy, deformation energy}

\maketitle

\section{Introduction}\label{intro}
The suggested more that 50 years ago macroscopic-microscopic approach \cite{swiat63,strut66} up to now is one of the most effective method for the calculations of quasistatic properties of atomic nuclei like ground state masses and deformations, the potential energy surface, the fission barriers and so on. In this method the energy of nucleus is represented as the sum of macroscopic and microscopic terms. The macroscopic part is often calculated within the liquid-drop model or finite range droplet model and for the  microscopic part the Strutinsky shell correction method \cite{strut67,strut68,brdapa} is used. At zero excitation energy these models allow for the very fast calculation of the energy of nucleus for any  shape . The ground state masses and deformations were calculated by macroscopic-microscopic method and tabulated in \cite{moller95,moller2000} for few thousands of atomic nuclei.

In nuclear reactions, however, the compound nuclei are formed at some excitations. Though, the generalization of the shell corrections to finite excitation (temperature) is quite straightforward, the calculations of the temperature dependence of the shell corrections is quite time consuming. As it was noted in \cite{ranmol} "Although, it would, in principle, straightforward to recalculate the shell+plus+pairing correction for specified finite temperatures, this would, in practice, be a rather formidable task if carried out for all of the over five million shapes of more than five thousand nuclei for which the original tabulation \cite{moller2000} was performed".

Instead, in many calculations the approximation 
\bel{ignat}
\delta E(E^*)=\delta E(E^*=0)\exp{(-E^*/E_d)}
\end{equation}
for the dependence of the shell correction on the excitation energy suggested in \cite{ignat75} for the phenomenological description of energy dependence of the level density parameter is used.
It was pointed out in \cite{ignat75} that approximation \req{ignat} is based on the Fermi-gas relations and does not account for the pairing correlations. The role of the pairing correlation and collective effects in the systematics of the level density of nuclei was considered in later work \cite{ignat79}. 

Still, the approximation \req{ignat} is used in many theoretical models both with and without account of pairing.
Together with the shell corrections at zero excitation energy tabulated in \cite{moller95,moller2000} the ansatz \req{ignat} offers a very simple way to account for the temperature dependence of the shell corrections. The damping factor $E_d$ in \req{ignat} was found in \cite{ignat75} to be close to $E_d=20 \mev$. In practical calculations it is often used as a fitting parameter. Depending on the described experimental data and the used theoretical approach the value of  $E_d$ can vary from  $E_d=15 \mev$ \cite{capote} to $E_d=60 \mev$ \cite{ranmol}.

Another approximation for the temperature dependence of shell corrections used in the theory of nuclear fission, see, for example \cite{pomo2006}, is the functional form for the shell correction to free energy $\delta F(T)$ suggested in \cite{bohrmo2} for the closed shell nuclei,
\bel{phitau}
\delta F(T)=\delta F(0)\Phi_{BM}(T),\,\Phi_{BM}(T)\equiv\tau/\sinh(\tau),
\end{equation}
where $\tau\equiv{2\pi^2 T}/{\hbar\omega_{sh}}$ and the energy spacing between the shells $\hbar\omega_{sh}=41 \mev/A^{1/3}$. The approximation \req{phitau} does not contain any adjustable parameter. The only uncertainty comes from the level density parameter $a$ that appears in the Fermi-gas relation between the temperature and excitation energy, $E^*=aT^2$.

One of the puzzles set by experiments is the dependence of neutron multiplicity on the fragment mass number at low excitation energies \cite{muller84}, say below $E^*=10 \mev$.
At such excitation energies the shell and pairing effects are especially important and one should be sure that the shell corrections are calculated accurate enough.

\begin{figure}[ht]
\centering
\includegraphics[width=0.8\columnwidth]{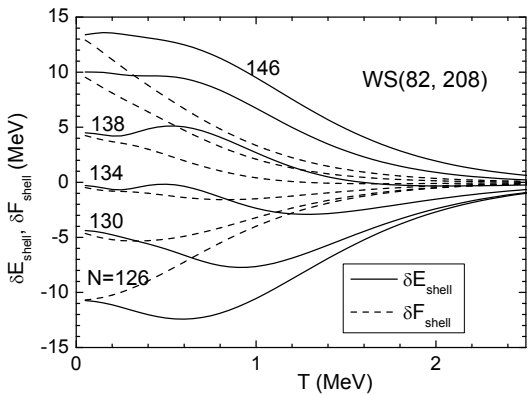}
\caption{The temperature dependence of the shell corrections to the energy $\delta E_{shell}$ (solid)  and free energy $\delta F_{shell}$ (dash) for the neutrons in the spherical Woods-Saxon potential of $^{208}$Pb, no pairing.}
\label{deltaef}
\end{figure}

Hard to believe, but in the last 50 years there were only few publications \cite{adeev72,adeev73,moretto72,aksel,bohrmo2,braque,civi82,civi83,ivahof} in which the temperature dependence of the shell correction was calculated directly.
The principal result of \cite{aksel,bohrmo2,ivahof} is reproduced in Fig.~\ref{deltaef}.
It turns out that only the shell correction to the free energy $\delta F_{shell}$ decays more or less exponentially. The temperature dependence of the shell correction to the  energy $\delta E_{shell}$ is more complicated.

That is why in present work we examine in details the temperature dependence of  the shell corrections including the shell corrections to the pairing energy and suggest the approximations to the energy and free energy that differ from a simple exponential decay.

In section \ref{nopair} the formal definitions of the shell corrections at $T\neq 0, \Delta=0$ are presented and some features are discussed. The simple approximations for the dependence of the shell corrections on the excitation energy are suggested. Sections \ref{pair} contains the analogous results for the case $\Delta\neq 0$. In section \ref{yields} we check the effect of derived here approximation for the shell correction on the mass distribution of fission fragments.
A short summary is given in section \ref{summa}.
\section{The shell corrections at $\Delta=0$}\label{nopair}
\begin{figure}[ht]
\centering
\includegraphics[width=0.99\columnwidth]{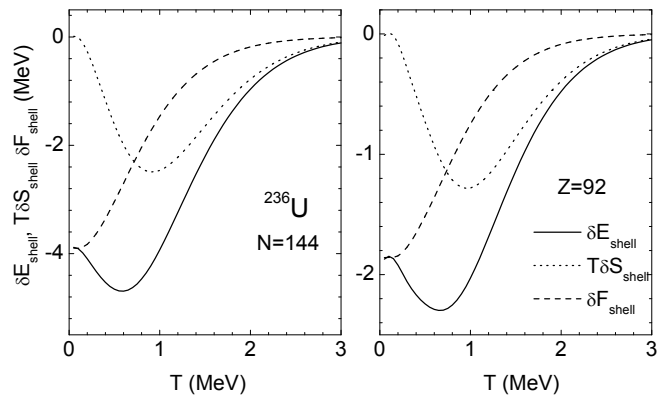}
\caption{The temperature dependence of the shell corrections to the energy \protect\req{deltaet} (solid), entropy \protect\req{deltast} (dot) and free energy \protect\req{deltaft} (dash) for the neutrons (left) and protons (right) at the ground state of $^{236}U$.}
\label{deltas}
\end{figure}

The shell correction to the energy of nucleus within the mean-field approximation is the difference between the sum $E_S$ of single-particle energies $\epsk$ of occupied states  and the averaged quantity $\widetilde E$,
\bel{deltae}
\delta E_{shell}=E_S-\widetilde E,
\end{equation}
where
\bel{eshell}
E_S=\sum_{occ.}\epsk =\int_{-\infty}^{\epsilon_F} e g_S(e) de,\,\,g_S(e)\equiv 2\sum_k\delta(e-\epsk).
\end{equation}
The average part of energy is calculated by replacing in \req{eshell} the exact density of states $g_S(e)$ by the averaged quantity $\widetilde g(e)$,
\bel{gtilde}
\widetilde g(e)=
\frac{1}{\gamma}\int_{-\infty}^{\infty}f\left(\frac{e-e^{\pr}}{\gamma}\right)g_S(e^{\pr})de^{\pr}=\frac{2}{\gamma}\sum_k f\left(\frac{\epsk-e}{\gamma}\right),
\end{equation}
\bel{etilde}
\widetilde E=\int_{-\infty}^{\tilde\mu} e \widetilde g(e)\,de\,,
\end{equation}where $f(x)$ is the so-called Strutinsky smoothing function
\bel{fx}
f(x)=\frac{e^{-x^2}}{\sqrt{\pi}}\sum_{n=0, 2 ...}^M a_n H_n(x),a_0=1, a_{n+2}=\frac{-a_n}{n+2}.
\end{equation}
The generalization of Eqs.\req{deltae}-\req{etilde} to finite temperature is quite straightforward. For the energy $E(T)$ of system of independent particles at finite temperature one has
\bel{est}
E(T)=2\sum_k\epsk n_k^T, \, \text{with}\, n_k^T=\frac{1}{1+e^{(\epsk-\mu)/T}}.
\end{equation}
The averaged energy $\widetilde E(T)$ is defined by replacing the sum in \req{est} by the integral with the smoothed density of states $\widetilde g(e)$
\bel{etavr}
\widetilde E(T)=\int_{-\infty}^{\infty}de \widetilde g(e)e n_e^T,
\end{equation}
with $n_e^T\equiv 1/[1+e^{(e-\tilde\mu)/T}]$. 
The chemical potentials $\mu$ and $\tilde\mu$ in \req{est}-\req{etavr} are defined by the particle conservation condition,
\bel{mues}
2\sum_k n_k^T=\int_{-\infty}^{\infty}de \widetilde g(e) n_e^T = N.
\end{equation}
The integrals in \req{etavr}-\req{mues} should be calculated numerically. The details are given in the Appendix A.
The shell correction to the energy at finite temperature is then
\bel{deltaet}
\delta E_{shell}(T)=E(T)-\widetilde E(T).
\end{equation}
Another quantity of interest is the shell correction to free energy
\bel{deltaft}
\delta F_{shell}(T)=\delta E_{shell}(T) -T \delta S_{shell}(T),
\end{equation}
(the driving force in Langevin equations \cite{abe-sun} is given by the derivative of free energy with respect to deformation at fixed temperature).
For the entropy we use the standard definition of $S(T)$ for the system of independent particles
\bel{st}
S(T)=-2\sum_k [n_k^T \log n_k^T +(1-n_k^T)\log(1-n_k^T)].
\end{equation}
The average part of $S(T)$ is defined in an analogous way by the replacing the sum in \req{st} by the integral
\bel{stavr}
\widetilde S(T)=-\int_{-\infty}^{\infty}de \widetilde g(e) [n_e^T \log n_e^T +(1-n_e^T)\log(1-n_e^T)].
\end{equation}
And the shell correction to the entropy is the difference between \req{st} and \req{stavr},
\bel{deltast}
\delta S_{shell}(T)=S(T)-\widetilde S(T).
\end{equation}
The calculated shell corrections to the energy, entropy and free energy are shown in Fig.~\ref{deltas}. The calculations are carried out with the Woods-Saxon potential \cite{pash71,pash88} for the ground state of $^{236}\rm{U}$ which is the most important for the applications related to the atomic energy problems. The parameters of the potential are taken from \cite{pash08}.

First of all, one notice the non-monotonous dependence of $\delta E_{shell}$ on temperature for protons. The shell correction $\delta E_{shell}$ grows (in absolute value) up to $T\approx 0.5 \mev$ (i.e. shell effects become stronger) and then falls down. Such behaviour was noticed already in \cite{aksel,bohrmo2}. The $\delta E_{shell}$ is the difference between $E(T)$ and $\widetilde E(T)$. Both quantities decrease with temperature but with different speed at small $T$. The dependence of $E(T)$ on $T$ at small $T$ is sensitive to the position of individual single-particle levels near the Fermi energy. Depending on whether the local density of these states is larger or smaller than the average, the $\delta E_{shell}$ will grow or decrease with $T$ at small $T$.

It is evident from Fig.~\ref{deltas} that the temperature dependence of $\delta E_{shell}$ differs substantially from the approximation \req{ignat}.
The shell correction to the free energy $\delta F_{shell}$ looks, on contrary, very similar to \req{ignat}.

In order not to be bound by the peculiarities of the ground state shape, we have calculated the ratio  of total (neutrons plus protons) shell correction to free energy to its value at $T=0$, $\delta F_{shell}(T)/\delta F_{shell}(0)$, averaged over more than 1000 points in the deformation space. More precisely, we used three dimensional mesh with the grid points in $0\leq\alpha\leq 1, \Delta\alpha=0.1$, $-0.5\leq\alpha_1\leq 0.5, \Delta\alpha_1=0.1$, $-0.5\leq\alpha_4\leq 0.5, \Delta\alpha_4=0.1$. The deformation parameters $\alpha, \alpha_1$ and $\alpha_4$ of Cassini shape parametrization describe the total elongation of nucleus, the mass asymmetry and the neck radius, i.e. the main fission degrees of freedom, see \cite{pash71,pash88}. The $\alpha=0$ corresponds to spherical shape, $\alpha=1$ corresponds to the shape with zero neck radius.

At $T=0$ the shell correction $\delta F(0)$ may have different sign at different deformation point and $\delta F$ averaged in deformation space has not much sense. The ratio  $\delta F(T)/\delta F(0)$ at each deformation point is equal to one for $T=0$ and then decreases somehow with growing $T$. Thus, the sum of $\delta F(T)/\delta F(0)$ over many deformation points gives information on the average variation of $\delta F(T)$ with the temperature.

\begin{figure}[ht]
\centering
\includegraphics[width=0.9\columnwidth]{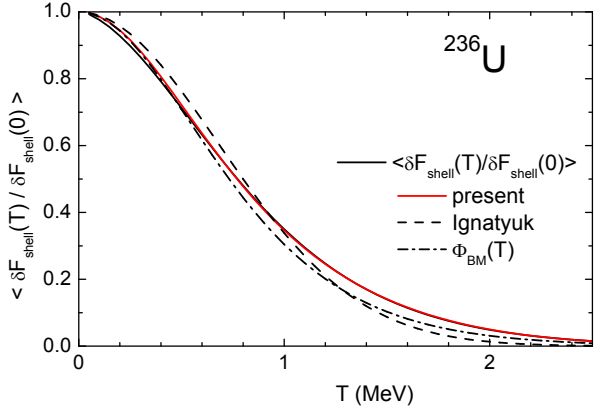}
\caption{The temperature dependence of the averaged in deformation ratio $\delta F_{shell}(T)/\delta F_{shell}(0)$ for $^{236}$U (black solid) and parameterisations  \req{ignat} (dash), \req{phitau} (dash-dot) and \req{PhiIvn} (red solid).}
\label{Phis236}
\end{figure}

In Fig.~\ref{Phis236} we compare the averaged in deformation ratio $\delta F_{shell}(T)/\delta F_{shell}(0)$ with the parameterisations of \cite{ignat75}.

One can see that approximation \req{ignat} with $E_d=20 \mev$ is rather close to the calculated average value $\langle\delta F(T) / \delta F(0)\rangle$.
The approximation \req{phitau} is slightly better, even without adjustable parameters. The temperature in \req{phitau} was related to the excitation energy by $E^*=\tilde a T^2$, with $\tilde a$ given by Eq. \req{aat} below.

For more accurate approximation of $\langle\delta F(T) / \delta F(0)\rangle$ we have fitted it by the two-parametric curve, similar to that suggested in \cite{ranmol},
\bel{PhiIvn}
\Phi(E^*)=(e^{-E_1/E_0}-1)/(e^{(E^*-E_1)/E_0}-1),
\end{equation}
see red curve in Fig.~\ref{Phis236}.
In case of $^{236}{\rm U}$ the fit leads to the values $E_0=42.28 \mev$, $E_1=-18.54 \mev$. The original quantity $\langle\delta F(T) / \delta F(0)\rangle$ (black curve in Fig.~\ref{Phis236}) and the fit \req{PhiIvn} are almost identical.

So, in cases when the pairing can be neglected, the temperature dependence of the shell correction to the free energy can be accurately approximated by
\bel{deltaf-fit}
\delta F_{shell}(E^*)=\delta F_{shell}(0)\Phi(E^*),
\end{equation}
with $\Phi(E^*)$ given by \req{PhiIvn}.

The input quantity in the calculations is the temperature, the parameter that appears in the Fermi functions for the occupation numbers. For given temperature one can calculate the excitation energy, $E^*=E(T)-E(0)$ and plot various quantities both as functions of temperature of excitation energy.

\begin{figure}[ht]
\centering
\includegraphics[width=0.9\columnwidth]{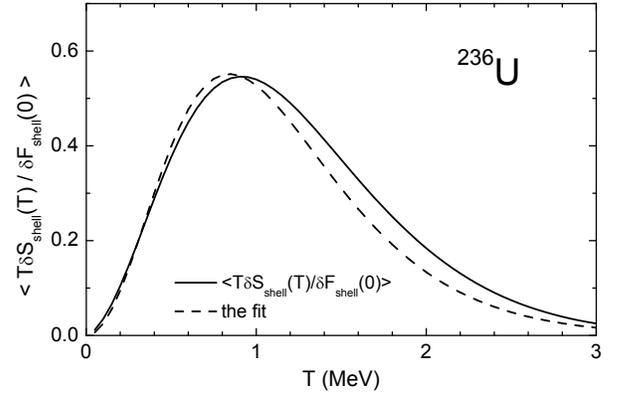}
\caption{The average value of $\langle T\delta S(T))/\delta F(0)\rangle$ (solid) and the fit \protect\req{PhiSBM} (dot-dash).}
\label{fitdels}
\end{figure}

For the shell correction to the energy we would need the similar approximation for the temperature dependence of the shell correction to the entropy \req{deltast}. For this purpose we have calculated the averaged in deformation ratio $T\delta S_{shell}(T)/\delta F_{shell}(0)$ and fitted it by the functional form derived in \cite{bohrmo2} for the closed shell nuclei
\bel{PhiSBM}
T\delta S_{shell}(T)/\delta F_{shell}(0)=T\delta S_0[\tau\coth(\tau)-1]/\sinh(\tau),
\end{equation}
where $\tau\equiv{2\pi^2 T}/{\hbar\omega_{sh}}$ and $\hbar\omega_{sh}$ being the energy spacing between the shells, $\hbar\omega_{sh}=41 \mev/A^{1/3}$.
For the $\delta S_0$ we obtained in this way the value $\delta S_0=2.5 \mev^{-1}$.
The comparison of the average value of $\langle T\delta S_{shell}(T)/\delta F_{shell}(0)\rangle$ and the fit \req{PhiSBM} is shown in Fig.~\ref{fitdels}. Note, that the fit \req{PhiSBM} contains only one fitted parameter $\delta S_0$.
In principle, as it follows from \cite{bohrmo2}, the quantity $\delta S_0$ depends also on $T$. For simplicity we have neglected this dependence. That is why for large values of $T$ the calculated values of $\langle T\delta S_{shell}(T)/\delta F_{shell}(0)\rangle$, and the fit, differ somewhat from each other.

Putting together the approximations for $\delta F(T)$ and $\delta S(T)$ the approximation for $\delta E(T)$ takes the form
\bel{deltae-fit}
\delta E_{shell}(T)=\left[\Phi(E^*)+T \delta S_0\frac{\tau\coth(\tau)-1}{\sinh(\tau)}\right]\delta F_{shell}(0).
\end{equation}

\begin{figure}[ht]
\centering
\includegraphics[width=0.9\columnwidth]{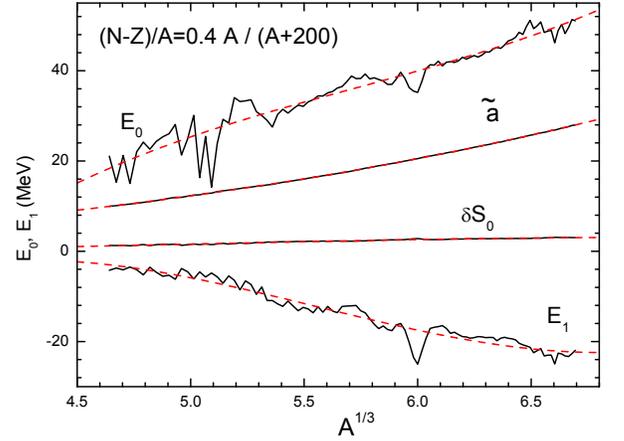}
\caption{The dependence of parameters $E_0$, $E_1$ (in $\mev$) and $\delta S_0, \tilde a$ (in $\mev^{-1}$) on the mass number A.}
\label{averag}
\end{figure}

In order to establish the dependence of parameters $E_0$, $E_1$ and $\delta S_0$ and the level density parameter $\tilde a$ on the mass number $A$ we have carried out the fit of $\langle\delta F_{shell}(T) / \delta F(_{shell}0)\rangle$ and
$\langle T\delta S_{shell}(T))/\delta F_{shell}(0)\rangle$ for the nuclei between $A=100$ and $A=300$ along the  beta-stability line \cite{swiat74}. The brackets $\langle ... \rangle$ mean here the averaging in deformation as explained above. The obtained results for the $A$-dependence of $E_0$, $E_1$ and $\delta S_0$, see Fig.~\ref{averag}, were fitted by the polynomial in $A^{1/3}$ (dash lines). 

In this way we got the approximations:
\belar{constants}
E_0&\approx& (-467+236 A^{1/3}-38.6 A^{2/3}+2.24 A)\mev  \nonumber\\
E_1&\approx&(-391+230 A^{1/3}-43.6 A^{2/3}+2.62 A)\mev    \nonumber\\
\delta S_0&\approx&8.65-6.29 A^{1/3}+1.45 A^{2/3}-0.0949 A
\end{eqnarray}
In the same way we have estimated the averaged in deformation value of the level density parameter
\bel{aat}
\tilde a=\frac{\pi^2}{6}\tilde g (\tilde \mu),\quad\text{with}\quad\tilde g (\tilde \mu)=\frac{2}{\gamma}\sum_k f\left(\frac{\epsk-\tilde\mu}{\gamma}\right)\,.
\end{equation}
\bel{denspar}
\tilde a \approx (0.0984 A-0.253 A^{2/3}+2.07 A^{1/3}-4.04)/\mev.
\end{equation}
In principle, the density of levels \req{aat} depends on the shape of nucleus. In Langevin calculations the shape of nucleus varies in a very broad region of elongation and mass asymmetry. Since we use parameter $\tilde a$ in Langevin calculations, in Fig.~\ref{averag} we show the value averaged over the whole region of deformations. The level density parameter for the ground state may differ from the approximation \req{denspar}.

\begin{figure}[ht]
\centering
\includegraphics[width=0.99\columnwidth]{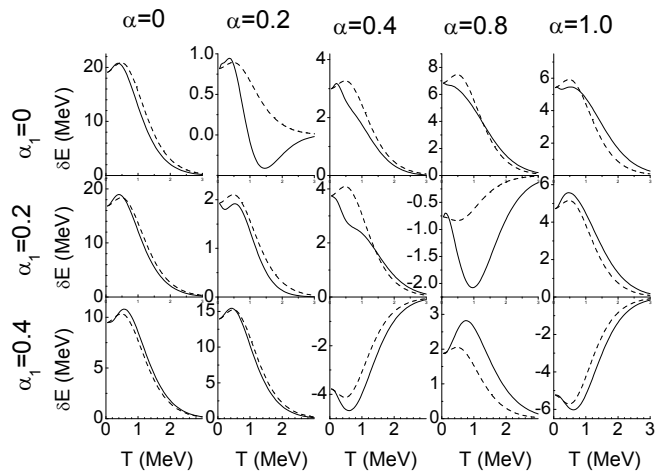}
\caption{The temperature dependence of the calculated shell corrections to the energy \protect\req{deltaet} (solid) and the approximation \protect\req{deltae-fit} (dash) for $^{236}U$.}
\label{deltaet-fit}
\end{figure}

The comparison of calculated shell corrections to the energy \req{deltaet} and the approximation \req{deltae-fit} for different points in the deformation space of $^{236}$U is shown in Fig.~\ref{deltaet-fit}.

One can see that approximation \req{deltae-fit} on average correctly reproduces the temperature dependence of $\delta E(T)$. Both grows with the temperature at small temperature, reach the maximum at approximately the same $T$ and are rather close to each other in region of larger $T$, say for $T\ge 1\mev$. The $\delta E(T)$ and the fit differ substantially only at the deformation points where the shell correction is very small, $\delta E \approx 1\sim 2 \mev$, and changes its sign with raise of temperature. It is clear that such dependence can not be described by the simple approximation. In such cases, $\delta E(T)$ should be calculated directly, if necessary (it is relatively small and should not be important).
\section{The shell correction to the pairing energy at finite temperature}\label{pair}
Like in \cite{aksel,bohrmo2,brdapa} and many other works, we account for the pairing interaction in Bardeen-Cooper-Schriffer (BSC) approximation \cite{bardeen}.
For the energy of independent quasi-particles at finite temperature one has \cite{ivahof},
\belar{ebcst}
E_{BCS}(T)=2\sum_{k=k_1}^{k_2}\epsk n_k^{\Delta,T}-\frac{\Delta^2}{G},\quad\text{with}\nonumber\\
n_k^{\Delta,T}\equiv \frac{1}{2}\left(1-\frac{\epsk-\lambda}{E_k}\tanh\frac{E_k}{2T}\right),
\end{eqnarray}
where $G$ is the strength of the pairings interaction, $k_1$ and $k_2$ - the limits of the so called pairing window, $\Delta$ is the pairing gap and $E_k$ are the quasi-particle energies,
\bel{eqp}
E_k=\sqrt{(\epsk-\lambda)^2+\Delta^2}.
\end{equation}

For given particle number $N$ and the pairing strength $G$ the chemical potential $\lambda$ and the pairing gap $\Delta$ are found from the particle number conservation and the gap equation,
\bel{nbcst}
2\sum_{k=k_1}^{k_2}n_k^{\Delta,T}=N-2k_1+2\,,\,
\sum_{k=k_1}^{k_2}\frac{1}{E_k}\tanh\frac{E_k}{2T}=\frac{2}{G}\,.
\end{equation}
The summation in finite limits in \req{ebcst}, \req{nbcst} is dictated by the BCS pairing approximation. The constant
pairing strength $G$ can only be assumed in a finite interval around  the  Fermi energy,  otherwise the summation in gap
equation \req{nbcst} would diverge.

For the entropy one has the expression analogous to \req{st},
\bel{sbcst}
S(T)=-2\sum_{k=k_1}^{k_2}[n_k^{\Delta,T} \log n_k^{\Delta,T} +(1-n_k^{\Delta,T})\log(1-n_k^{\Delta,T})]
\end{equation}
The pairing energy is defined then by the difference between \req{ebcst} and the energy of independent particles within the pairing gap,
\bel{epairt}
E_{pair}(T)=E_{BCS}(T)-2\sum_{k=k_1}^{k_2} \epsk n_k^T\,.
\end{equation}
Similar, the pairing contribution to the entropy is
\belar{spairt}
S_{pair}(T)=2\sum_{k=k_1}^{k_2} [n_k^T \log n_k^T +(1-n_k^T)\log(1-n_k^T)]\nonumber\\
-2\sum_{k=k_1}^{k_2}[n_k^{\Delta,T} \log n_k^{\Delta,T} +(1-n_k^{\Delta,T})\log(1-n_k^{\Delta,T})].
\end{eqnarray}
The average counterparts of $E_{pair}$ and $S_{pair}$ are defined by neglecting the shell effects in \req{epairt}-\req{spairt}, i.e. by replacing the sum over quantal states $\vert k\rangle$ by the integrals with the average density of single-particle states \req{gtilde} defined in terms of Strutinsky smoothing,
\belar{epairavr}
\widetilde E_{pair}(T)&=&\frac{1}{2}\int_{\widetilde\lambda-s}^{\widetilde\lambda+s} de\tilde g(e) e\left(1-\frac{e-\widetilde\lambda}{E}\tanh\frac{E}{2T}\right)\nonumber\\
&-&\frac{\widetilde\Delta^2}{G}-\int_{\mu-s}^{\mu+s} de\tilde g(e) e n^T(e)\,,
\end{eqnarray}
and
\belar{spairavr}
\widetilde S_{pair}(T)&=&\int_{\widetilde\lambda-s}^{\widetilde\lambda+s} de\tilde g(e)  \left[\log(1+e^{-E/T})+\frac{E/T}{1+e^{E/T}}\right]\nonumber\\
&+&\int_{\mu-s}^{\mu+s} de\tilde g(e)[n^T(e) \log n^T(e) \nonumber\\
&+&(1-n^T(e))\log(1-n_T(e))],
\end{eqnarray}
with $n^T(e)\equiv 1/[1+e^{(e-\mu)/T}]$, $E\equiv \sqrt{(e-\widetilde\lambda)^2+\widetilde\Delta^2}$.
The chemical potential $\widetilde\lambda$ and the pairing gap $\widetilde\Delta$ for the system without shell effects are defined by the analog of \req{nbcst},
\belar{nbcstavr}
N-2k_1+2&=&\frac{1}{2}\int_{\widetilde\lambda-s}^{\widetilde\lambda+s} de\tilde g(e)\left(1-\frac{e-\widetilde\lambda}{E}\tanh\frac{E}{2T}\right),\,\nonumber\\
\frac{2}{G}&=&\frac{1}{2}\int_{\widetilde\lambda-s}^{\widetilde\lambda+s} de\tilde g(e)
\frac{1}{E}\tanh\frac{E}{2T}.
\end{eqnarray}
As the temperature increases, both pairing gaps $\Delta$ and $\widetilde\Delta$ decrease until they vanish at some critical temperature  $T_{crit}\approx 0.5 \mev$, which is somewhat different for $\Delta$ and $\widetilde\Delta$.
\begin{figure}[ht]
\centering
\includegraphics[width=0.99\columnwidth]{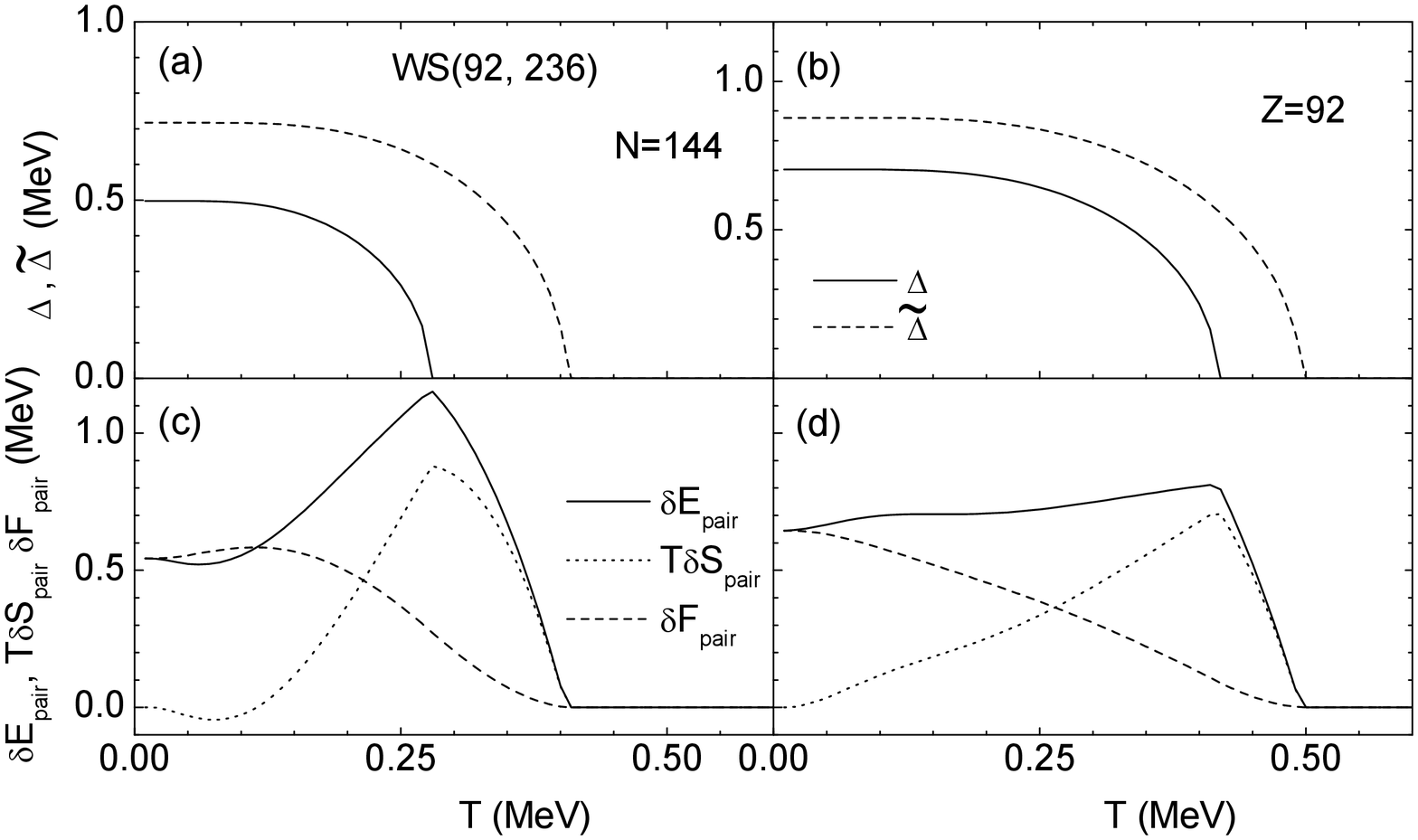}
\caption{Top: The temperature dependence of the pairing gaps $\Delta$ (solid) and $\widetilde\Delta$ (dot), defined by \protect\req{nbcst} and \protect\req{nbcstavr}, correspondingly. Bottom: The shell corrections to the pairing energy $\delta E_{pair}$ (solid), pairing free energy $\delta F_{pair}$ (dash) and the pairing entropy $\delta S_{pair}$ (dot) for neutrons and protons at the ground state of $^{236}$U.}
\label{deltaspair}
\end{figure}

The dependence of $\Delta$ and $\widetilde\Delta$ on $T$ for $^{236}$U is shown in  Figs.~\ref{deltaspair}(a, b). The Figs.~\ref{deltaspair}(c, d) show the calculated shell corrections to the
pairing energy $\delta E_{pair}(T)=E_{pair}(T)-\widetilde E_{pair}(T)$ , the pairing entropy
$\delta S_{pair}(T)=S_{pair}(T)-\widetilde S_{pair}(T)$ and to the pairing free energy $\delta F_{pair}(T)=  \delta E_{pair}(T)+T \delta S_{pair}(T)$.
Like in no-paring case, the shell correction $\delta E_{pair}$ first grows (in absolute value) as the temperature increases and then tends to zero. Unlike the no-paring case all the pairing shell corrections vanish at $T\approx 0.5 \mev$, when $\Delta$ turns into zero. Thus, at $T\ge T_{crit}$ only the shell corrections of independent particles remain.

\begin{figure}[hb]
\centering
\includegraphics[width=0.8\columnwidth]{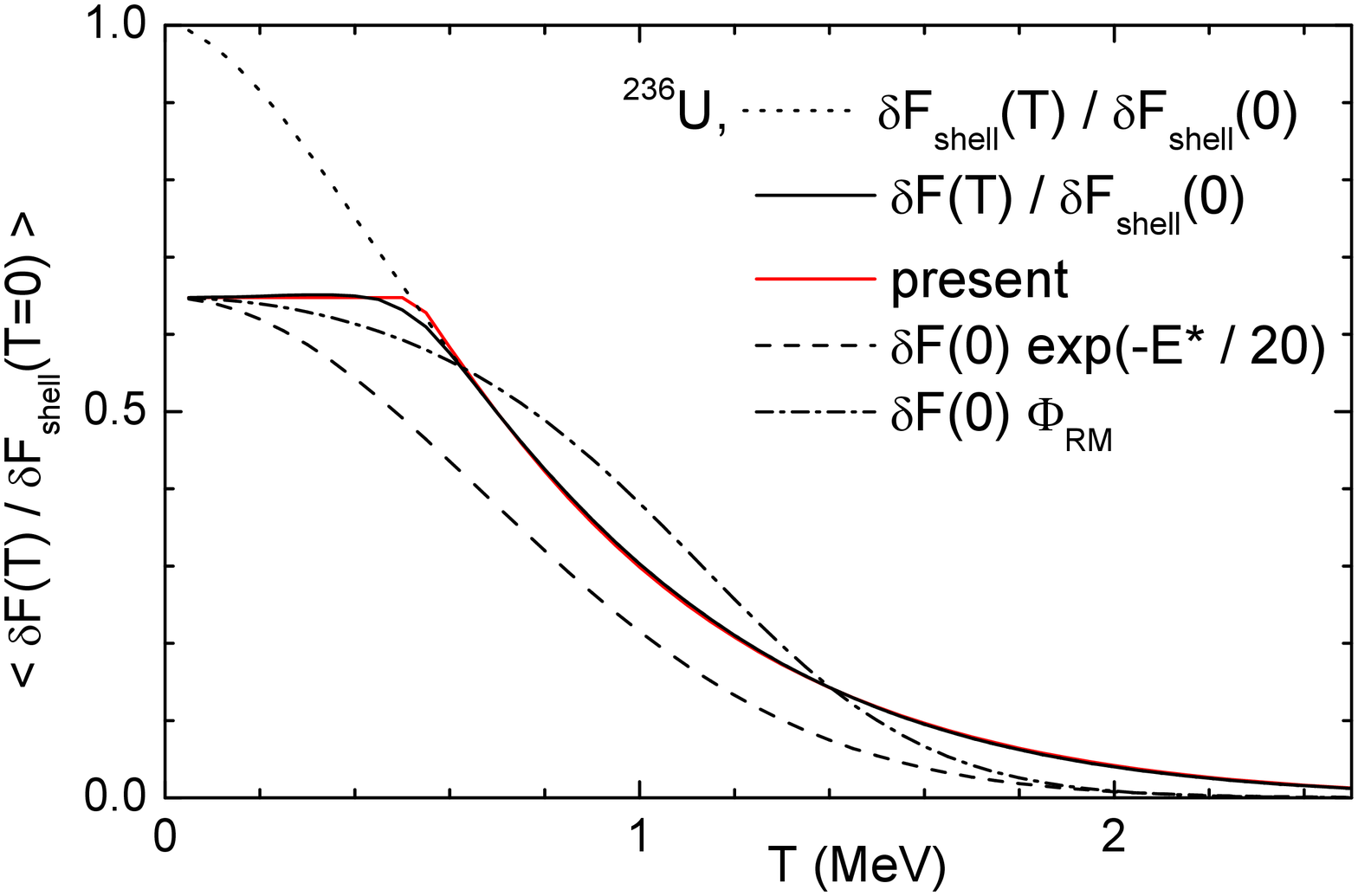}
\caption{The averaged in deformation temperature dependence of $\delta F_{shell}$ (dot), $\delta F_{shell}+\delta F_{pair}$ (black) and the approximation \protect\req{ignat} (dash), \protect\req{Phiran} (dot-dash) and \protect\req{delffit} (red).}
\label{fit_ivn}
\end{figure}

The averaged in deformation space total shell corrections (for protons plus neutrons) to the free energy $\langle (\delta F_{shell}(T)+\delta F_{pair}(T))/\delta F_{shell}(0)\rangle$ and the energy $\langle (\delta E_{shell}(T)+\delta E_{pair}(T))/\delta E_{shell}(0)\rangle$ are shown in Fig.~\ref{fit_ivn}.

It turns out that the shell and pairing corrections to free energy decrease with growing temperature almost with the same speed, so that below $T_{crit}$ the sum $\delta F_{shell}(T)+\delta F_{pair}(T)$ is almost constant.
 Thus, a good approximation to $\delta F(T)\equiv\delta F_{shell}(T)+\delta F_{pair}(T)$ would be a constant, equal to $\delta F(0)\equiv\delta F_{shell}(0)+\delta F_{pair}(0)$ below critical temperature and the approximation \req{PhiIvn} for $\delta F_{shell}$ above the critical temperature,
\belar{delffit}
\delta F(E^*)=\left\{
\begin{array}{rcl}
&\delta F(0),\,\text{if}\,\vert\delta F_{shell}(0)\Phi(E^*)\vert\ge \vert\delta F(0)\vert\,,\\
&\delta F_{shell}(0)\Phi(E^*),\\
&\text{if}\,\vert\delta F_{shell}(0)\Phi(E^*)\vert\leq \vert\delta F(0)\vert\,.
\end{array} \right.
\end{eqnarray}
The approximation \req{delffit} is shown in Fig.~\ref{fit_ivn} by the dash line. It almost coincides with the calculated shell correction $\delta F_{shell}(T)+\delta F_{pair}(T)$. For comparison we show also the approximations \req{ignat} and  the approximation of Randrup and M\"oller \cite{ranmol},
\bel{Phiran}
\Phi_{RM}(E^*)=(e^{-E_1/E_0}+1)/(e^{(E^*-E_1)/E_0}+1),
\end{equation}
with $E_0=15 \mev$, $E_1=20 \mev$.
As one can see, these approximations deviate substantially from the calculated shell corrections both at small and large temperatures.

The approximation \req{delffit} is well in line with the results of \cite{ignat79}, where it was shown that for the accurate desciption of the level density of nuclei below critical temperature one should use the temperature independent value of the level density parameter, $a=a(T_{crit})$.

\begin{figure}[ht]
\centering
\includegraphics[width=0.99\columnwidth]{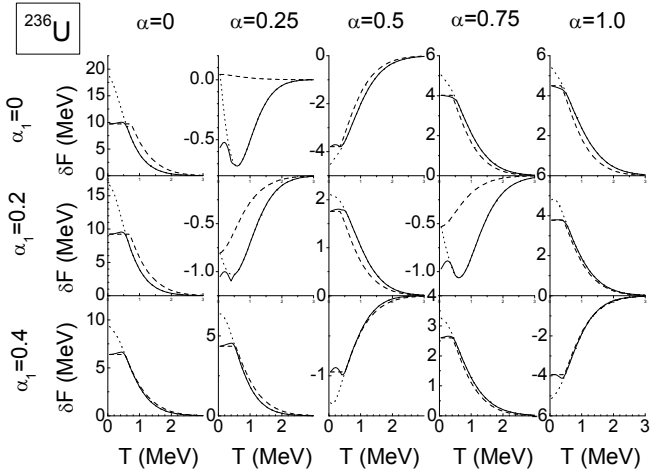}
\caption{The temperature dependence of the calculated shell corrections to the free energy $\delta F_{shell}$ \protect\req{deltaft} (dot) $\delta F_{shell}+\delta F_{pair}$ (solid) and the approximation \protect\req{delffit} (dash) for $^{236}$U.}
\label{delf236}
\end{figure}

In order to check how good is the approximation \req{delffit} we have calculated the dependence of $\delta F_{shell}(T)+\delta F_{pair}(T)$ on temperature for many points in the deformation space of $^{236}$U which differ in elongation and mass asymmetry and compared with approximation \req{delffit}. The results are shown in Fig.~\ref{delf236}.

One can see that in cases when $\delta F(T)$ is rather large, say larger that 2 MeV, the approximation \req{delffit} is rather close to the calculated $\delta F(T)$. The substantial deviations are seen only in cases when $\delta F(T)$ is of the order of $1\sim 2 \mev$.
Eventually, one should not expect better accuracy from \req{delffit}. It it meant to describe only the average dependence of $\delta F(T)$ on the temperature. At each particular deformation point there should be individual deviations from the average trend.

The approximation of $\delta E(T)$ is somewhat more difficult, mainly because it is not easy to fit the temperature dependence of $\delta E_{pair}(T)$. The reasonable simple approximation could be given by a Fermi function
\bel{delepfit}
\delta E_{pair}(T)\approx \delta E_{pair}(0)/[1+e^{(T-\widetilde T_{crit})/d}],
\end{equation}
with $d\approx 0.03 \mev$.
The $\delta E_{pair}(0)$ is easily calculated at zero temperature. The additional parameter that appears in \req{delepfit} is the critical temperature $\widetilde T_{crit}$ - the temperature at which the pairing effects vanish in the system with uniform distribution of single-particle states.
\begin{figure}[ht]
\centering
\includegraphics[width=0.9\columnwidth]{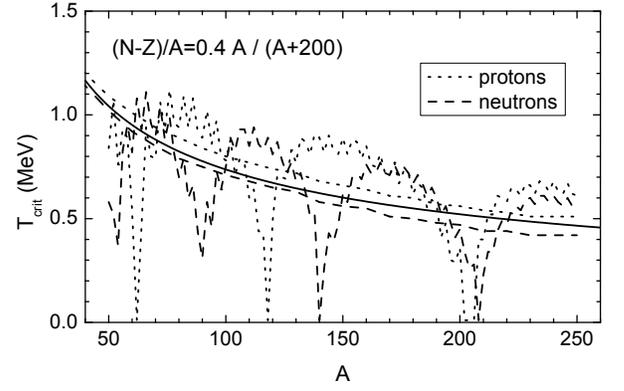}
\caption{The dependence of critical temperatures $T_{crit}$ and $\widetilde T_{crit}$ for neutrons (dash) and protons (dot) on the nucleus mass number $A$ along the beta-stability line. The solid line is the approximation \protect\req{tcrit}.}
\label{tcritica}
\end{figure}

To find an approximation for this quantity we have calculated both $T_{crit}$ and $\widetilde T_{crit}$ for neutrons and protons for the spherical nuclei with mass number $50\leq A \leq 250$ along the beta-stability line. The results are shown in Fig.~\ref{tcritica}.
As one could expect, the $T_{crit}(A)$ oscillates around the average value and turns into zero when the number of protons or neutrons is close to the magic number. The average over protons and neutrons value of $\widetilde T_{crit}(A)$ is nicely approximated by rather simple expression
\bel{tcrit}
\widetilde T_{crit}\approx {7.37 \mev} / {\sqrt{A}},
\end{equation}
(solid line in Fig.~\ref{tcritica}).

Putting together \req{deltae-fit} and \req{delepfit} one gets the approximation for the total shell correction $\delta E(T)=\delta E_{shell}(T)+\delta E_{pair}(T)$.
\begin{figure}[t]
\centering
\includegraphics[width=0.9\columnwidth]{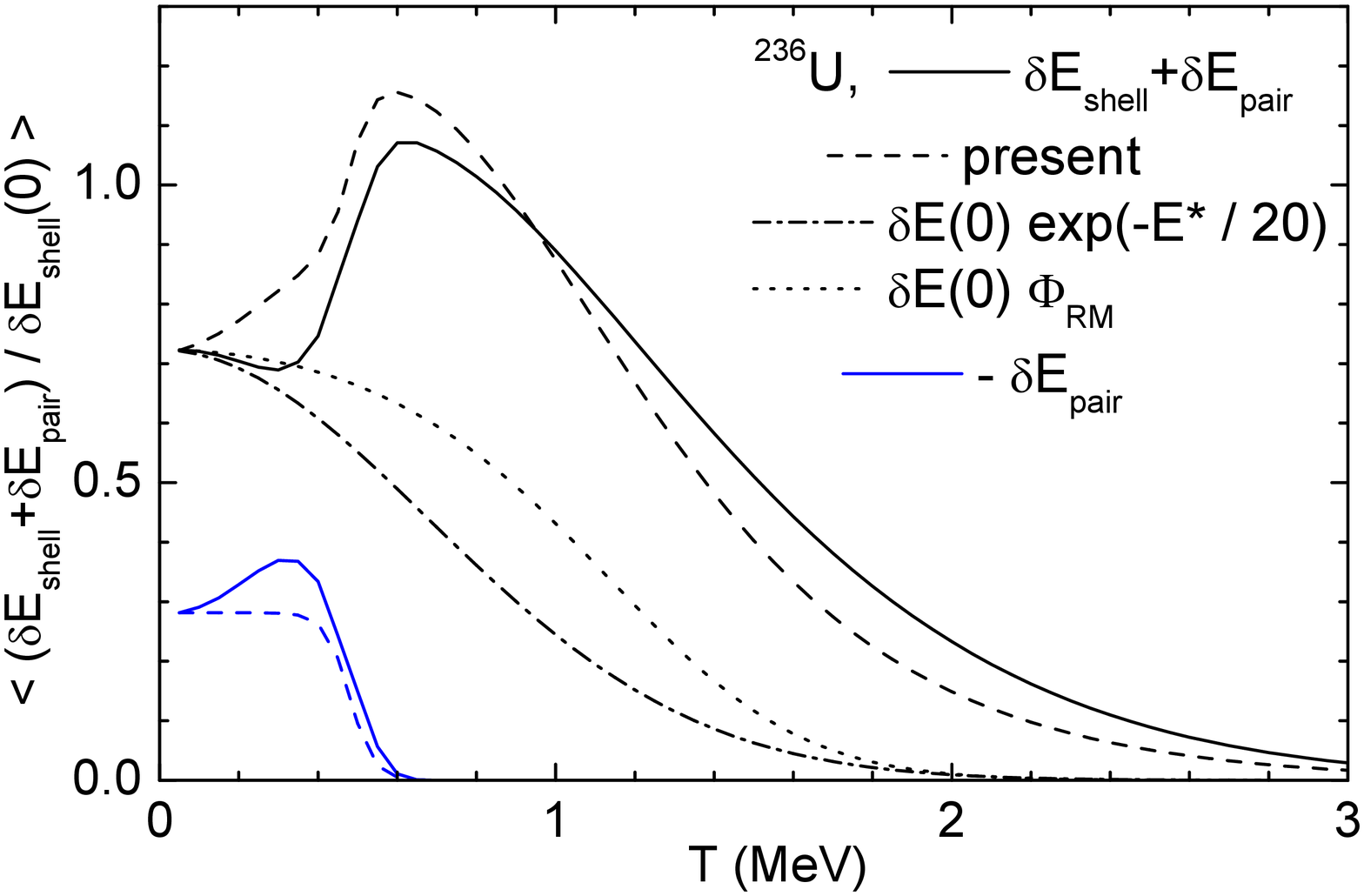}
\caption{The averaged in deformation temperature dependence of $-\delta E_{pair}$ (blue), $\delta E_{shell}+\delta E_{pair}$ (solid) and the approximations \protect\req{delepfit} (dash, blue) \protect\req{deltae-fit}, \protect\req{delepfit} (dash), \protect\req{ignat} (dot-dash), \protect\req{Phiran} (dot). }
\label{shelltot}
\end{figure}
Fig.~\ref{shelltot} shows the comparison of the averaged in deformation shell correction $\delta E_{shell}(T)+\delta E_{pair}(T)$ (solid) with the approximation \req{deltae-fit}, \req{delepfit} (dash). As one can see, the approximation \req{deltae-fit}, \req{delepfit} reproduces correctly the main features of the temperature dependence of the $\delta E(T)$: the raise at small temperatures, the position of maximum at $T\approx T_{crit}$ and the decay at higher temperatures. The main difference between $\delta E(T)$ and the approximation \req{deltae-fit}, \req{delepfit} comes from the not very accurate fit of the shell correction to the pairing energy (blue lines in Fig.~\ref{shelltot}).

\begin{figure}[h]
\centering
\includegraphics[width=0.99\columnwidth]{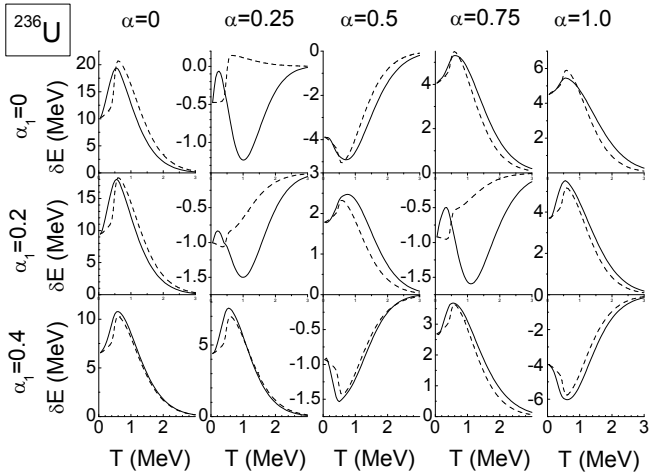}
\caption{The temperature dependence of the calculated total shell corrections to the energy $\delta E_{shell}+\delta E_{pair}$ (solid) and the approximation \protect\req{deltae-fit}, \protect\req{delepfit} (dash) for $^{236}$U.}
\label{deltaetot}
\end{figure}

The comparison of the calculated $\delta E(T)$ for some point in the deformation space of $^{236}$U with the approximation \req{deltae-fit}, \req{delepfit} is shown in Fig.~\ref{deltaetot}. Similar to the case of shell correction to free energy,
the approximation \req{deltae-fit},\req{delepfit} on average reproduces the temperature dependence of $\delta E(T)$ with the pairing effects included. Only when the shell correction is very small,  this approximation deviates substantially from the original quantity.
\section{The distributions of fission fragments}\label{yields}
The main conclusion from the investigation in present work is:

(I). The shell correction to free energy $\delta F$ does not decay exponentially at small excitation energies but is almost constant until the pairing vanishes and only then decays approximately exponentially.

This conclusion is confirmed by the results of \cite{ranmol}. The authors of \cite{ranmol} describe the charge distribution of fission fragments for a series of heavy nuclei at excitation energy $E^*= 11 \mev$ by means of Langevin equations for the overdamped motion. The driving force in Langevin equations is the derivative of {\it free} energy with respect to the deformation parameters. Here the shell correction to the free energy comes into play. From Fig.~2 of \cite{ranmol} one can see that the experimental charge distrubution can be reproduced only if the damping factor $E_d$ in \req{ignat} is very large, $E_d=60 \mev$ or even $E_d=\infty$. That means that at $E^*= 11 \mev$ the experimental results do not show the damping of shell effects, what is in agreement with our conclusion (I). 

Another conclusion of the present investigation concerns the dependence of the energy shell correction $\delta E$ on the temperature (excitation), see Fig.~\ref{deltaetot}.

(II). At $T\approx 1 \mev$ (the corresponding excitation energy is equal to $20\sim 30 \mev$, depending on the mass number) the energy shell correction $\delta E$ is (at least) as large as at $T=0$.

At present there are some indications that the shell effects in atomic nuclei are present at the excitation energies of the order of $50\sim 60 \mev$. In \cite{nishio2016,nishio2017} the mass distributions of fission fragments were measured of the nuclei populated by the multi-nucleon transfer channels in reactions
of $^{18}{\rm O}$ with isotopes of Th, U, Np, Pu. It is shown that even at $E^*=50\sim 60 \mev$ the mass distributions are clearly mass-asymmentric, what can be only due to the shell effects.

The accurate theoretical description of fusion-fission reactions is unfortunately very time-consuming. In addition to the usual Langevin calculations one has to evaluate and subtract the rotational energy and take into account the possibility of multi-chance fission. Such calculations would be a subject of a separate publication.

There are also a simpler experimental data. In \cite{ryzhov2011} the mass distributions of fission fragments in reactions $^{232}{\rm Th}$+n and $^{238}{\rm U}$+n were measured at the neutron energies $E_n=$32.8, 45.3 and 59.9 MeV. In all case the measured mass distributions are mass-asymmetric.
\begin{figure}[ht]
\centering
\includegraphics[width=0.9\columnwidth]{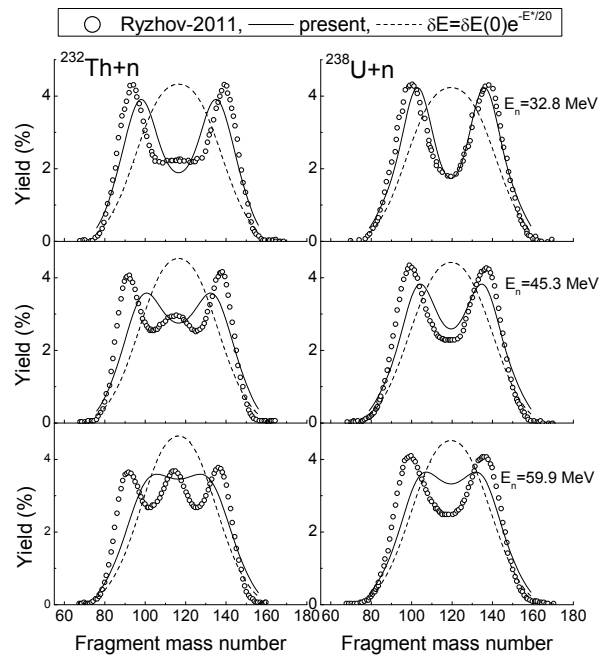}
\caption{The measured \protect\cite{ryzhov2011} (open circles) and calculated mass distributions of fission fragments in reactions $^{232}{\rm Th}$+n, $^{238}{\rm U}$+n. In calculated mass distributions \protect\req{yield_spm} the shell correction to the liquid drop energy was approximated by \protect\req{ignat} (dash) or \protect\req{deltae-fit}, \protect\req{delepfit} (solid). }
\label{yielda}
\end{figure}
The reactions with neutrons are somewhat simpler for the theoretical interpretation since in this case one has not to consider the rotational energy. Still, the application of dynamical approach would take a lot of time. In order to demonstrate the role of the new approximation \req{deltae-fit}, \req{delepfit} for $\delta E(E^*)$ we have estimated the mass distribution within the simpler approach - the scission point model \cite{spm,spm4,ivchar}, see also \cite{krapom}. In this model the mass distribution of fission fragments is defined only by the deformation energy at the scission line
\bel{yield_spm}
Y(\eta)\propto \sum_ie^{-E_{def}(\eta, \alpha_i)/T_{coll}}.
\end{equation}
Here $\eta$ is the mass asymmetry parameter and $T_{coll}$ is the width of distribution of deformation energy in the space of deformation parameters $\alpha_i$. The value of $T_{coll}$ was estimated in \cite{spm} to be close to $T_{coll}=$ 1 MeV. The $E_{def}$ in \req{yield_spm} is the deformation energy that includes both macroscopic part and the energy shell correction $\delta E=\delta E_{shell}+\delta E_{pair}$. 

We have carried out the calculation of the yield \req{yield_spm} for reactions $^{232}{\rm Th}$+n and $^{238}{\rm U}$+n  with the deformed Woods-Saxon potential \cite{pash71}. The shape of nuclear surface along the scission line was parameterized in terms of Cassini ovals, see \cite{pash71}, with 6 deformation parameters, $\alpha, \alpha_1, \alpha_2, \alpha_3, \alpha_4, \alpha_6$. The scission line was fixed by $\alpha=0.92$. The rest of deformation parameters was included in summation in \req{yield_spm} under the restriction that the mass asymmetry has a fixed value $\eta$.

The damping of shell effects with the excitation energy $E^*=E_n+B_n$ was taking into account by the approximation \req{ignat} or \req{deltae-fit}, \req{delepfit}. From Fig.~\ref{yielda} one can see that the mass distributions calculated with the approximation \req{deltae-fit}, \req{delepfit} are much closer to the experimental data compared with that obtained with \req{ignat}. The difference between calculated and experimental values seen in Fig.~\ref{yielda} is partly due to too simple approximations \req{yield_spm} and partly to the fact the $E^*=E_n+B_n$ is the excitation energy at the ground state. At scission the excitation energy could be very different. On one hand, the nucleus is getting more excited due to the dissipation of collective kinetic energy. On other - the excitation energy is taken away by the emitted particles and $\gamma$-rays. For more accurate description one would have to run very time consuming dynamical equations, what is beyond the scope of present work. In any case, from Fig.~\ref{yielda} the advantage of approximation \req{deltae-fit}, \req{delepfit} as compared with \req{ignat} is obvious.
\section{Summary}\label{summa}
We have calculated the temperature dependence of the shell corrections to the macroscopic nuclear energy directly starting from their formal definitions without any additional approximations.

We have demonstrated  that below critical temperature, where the pairing effects are important, both shell correction to energy and the shell correction to the free energy differ substantially from the popular approximation $\delta E(E^*)=\delta E(0)e^{-E^*/E_d}$. At small excitation energy the shell correction to the energy deviates from this approximation even when the pairing effects are absent.

It is shown that:
 
(I). The shell correction to free energy $\delta F$ does not decay exponentially at small excitation energies but is almost constant until the pairing vanishes and only then decays approximately exponentially.

(II). At $T\approx 1 \mev$ (the corresponding excitation energy is equal to $20\sim 30 \mev$, depending on the mass number) the energy shell correction $\delta E$ is (at least) as large as at $T=0$.

We have proposed the approximations for the shell corrections to the energy $\delta E$ and free energy $\delta F$ that reproduce rather accurately the average dependence of $\delta E$ and $\delta F$ on the temperature (excitation energy). These approximations rely on the quantities calculated at zero temperature $\delta E_{shell}(0)$ and $\delta E_{pair}(0)$ and few fitted constants.



\begin{acknowledgments}
This study comprises the results of "Research and development of an
innovative transmutation
system of LLFP by fast reactors", entrusted to the Tokyo Institute of
Technology by the Ministry of Education,
Culture, Sports, Science and Technology of Japan (MEXT).
We appreciate very much the usefull discussions with Prof. A.V. Ignatyuk. 
One of us (F. I.) would like to express his gratitude to the
Laboratory for Advanced Nuclear Energy
for the hospitality during his stay at Japan.
\end{acknowledgments}
\appendix
\section{The evaluation of averaged quantities}
\label{append}
In case $T=\Delta=0$ the averaged part of energy \req{etilde} can be transformed to the form
\belar{esmooth}
\widetilde E(0)=2\sum_k\epsk\tilde n_k- 2\gamma \sum_k\int_{x_k}^{\infty}xf(x)dx,\nonumber\\
\text{with}\quad x_k\equiv(\epsk-\tilde\mu)/\gamma,\quad \tilde n_k\equiv\int_{x_k}^{\infty}f(x)dx.
\end{eqnarray}
Since the smoothing function $f(x)$ is expressed in terms of Hermite polynomials the integrals in \req{esmooth} are calculated analytically using the recurrence relations between Hermite polynomials and their derivatives.

In case of non-zero temperature the averaged part of energy $\widetilde E(T)$ \req{etavr} is equal to
\belar{etsmooth}
&&\widetilde E(T)=\widetilde E(0)+\\
&&\int_0^{\infty}[(\tilde\mu+Tx)\tilde g(\tilde\mu+Tx)-(\tilde\mu-Tx)\tilde g(\tilde\mu-Tx)]n(x)dx,\nonumber
\end{eqnarray}
with $n(x)=1/(1+e^{x})$. At large $x$ the $n(x)$ is proportional to $e^{-x}$, so for the integral in \req{etsmooth} one can use the numerical methods for the integral of the type $\int_0^{\infty}f(x)e^{-x}dx$. By the same method one can calculate also the integral \req{stavr} for $\tilde S(T)$.

The calculation of the shell correction at finite temperature is much more time consuming compared with $T=0$ case. That is why we account for the pairing interaction here in the simplest BCS approximation. In this approximation the two additional parameters appear - the strength $G$ of pairing interaction and the size $2s$ of pairing window. The chemical potential $\lambda$ and the pairing gap $\Delta$ should be found from the pair of equations \req{nbcst}. Since $G$ depends sensitively on the nuclear region considered and on the details of pairing calculations it was suggested in \cite{moller92} to relate $G$ to the smooth pairing gap $\tilde\Delta$,
\bel{gapavr}
\frac{2}{G}=\widetilde g(\lambda){\rm ln}{[\sqrt{1+s^2/\widetilde\Delta^2}+s/\widetilde\Delta]}
\end{equation}
Following \cite{moller92} we used the following approximation for the average pairing gap $\widetilde\Delta$
\bel{ugap}
\widetilde\Delta=\left\{
\begin{array}{rl}
r\,e^{-tI^2+pI}/Z^{1/3}, \qquad\text{for protons},\,\,\,\, \\
r\,e^{-tI^2-pI} /N^{1/3}, \qquad\text{for neutrons},
\end{array} \right.
\end{equation}
with $r=5.72 MeV, p=0.118, t=8.12, I\equiv (N-Z)/A$. The $2s$ is the size of the pairing gap, $\lambda-s\leq\epsk\leq\lambda+s$, and the average density of states was assumed constant within the pairing window \cite{brdapa}, $\tilde g(e)=\tilde g(\lambda)$.
The $s$ is close to the spacing between the shells $\hbar\omega_{sh}$. In the code by V.Pashkevich $s$ was fixed by $s=1.1 \hbar\omega_{sh}$. The summation in \req{ebcst}-\req{spairt} is carried out over the states within the pairing window, $k_1=(N-\tilde g s)/2$,$k_2=(N+\tilde g s)/2$.
In the same approximation one gets for the smoothed  pairing energy
\bel{epairav}
\widetilde E_{pair}(T=0)=\widetilde g(\lambda)s^2[1-\sqrt{1+\widetilde\Delta^2/s^2}]\approx -\frac{1}{2}\widetilde g(\lambda)\widetilde\Delta^2
\end{equation}
At finite temperature the integrals in \req{epairavr}-\req{nbcstavr} should be calculated numerically. For this we used Simpson method.

\end{document}